# Towards robust detection of entangled two-photon absorption

[1,2,3,4]Raj Pandya*, [5]Patrick Cameron, [4]Chloé Vernière, [2,4]Baptiste Courme, [6]Sandrine Ithurria
[4]Alex Chin, [4]Emmanuel Lhuillier and [4]Hugo Defienne*

[1]Cavendish Laboratory, University of Cambridge, J.J. Thomson Avenue, CB3 0HE, Cambridge, UK
[2]Laboratoire Kastler Brossel, ENS-Université PSL, CNRS, Sorbonne Université, Collège de France, 24 rue Lhomond, 75005 Paris, France
[3]Department of Chemistry, University of Warwick, Coventry, CV4 7AL, UK
[4]Sorbonne Université, CNRS, Institut des NanoSciences de Paris, France
[5]School of Physics and Astronomy, University of Glasgow, Glasgow, UK
[6]Laboratoire de Physique et d'Etude des Matériaux, ESPCI-Paris, PSL Research University, Sorbonne Université, CNRS, Paris, France

*correspondence: raj.pandya@warwick.ac.uk, hugo.defienne@insp.upmc.fr

**Abstract**

Over the last 50 years entangled photon pairs have received attention for use in lowering the flux in two-photon absorption imaging and spectroscopy. Despite this, evidence for entangled two-photon absorption (ETPA) effects remain highly debated, especially at low-fluxes. Here, we structure the transverse spatial correlations of entangled photon pairs to evidence signs of ETPA at room-temperature in organic and inorganic chromophores, in the low-flux regime. We demonstrate our scheme to be robust to common artifacts that have previously hampered detection of ETPA such as linear absorption and background fluorescence, and show that ETPA scales with transverse correlation area and chromophore two-photon cross-sections. Our results present a step towards verifying ETPA and experimentally exploring entangled light-matter interactions.

**Introduction**

Two-photon absorption (TPA) is a key photophysical process exploited in a wide range of scenarios including biological imaging[1,2], (3D) lithography[3] and for characterizing the electronic structure of molecular and inorganic materials[4]. The problem is that TPA is extremely inefficient due to the low probability of two photons being absorbed nearly simultaneously (within the coherence time of the system, Δt; **Figure 1a**). To overcome this, applications of TPA currently rely on relatively high fluxes which may result in material damage. Entangled photon pairs have been proposed as a way to exploit TPA at low fluxes, in a so-called entangled two-photon absorption process (ETPA). Here the rate of two-photon absorption (TPA) is $R_{TPA} = \sigma_e \phi_{pair}$, *i.e.*, linear in flux, whereas in the classical case, $R_{TPA} = \delta_C \phi_{pair}^2$ which is quadratic in flux ($\sigma_e$ and $\delta_C$ are the entangled and classical two-photon absorption cross sections; $\phi_{pair}$ is the flux of photon pairs, entangled or otherwise)[5–11]. However, while theoretical works *e.g.*, by Landes *et al*.[12], show that TPA can be enhanced by several orders of magnitude using entangled photon pairs, they also find that the ETPA rates are too low for the process to be observable with many of the explored experimental configurations and chromophores. Indeed, the same authors even replicated a recent experiment which reported the observation of ETPA, without success[13,14]. Consequently, works claiming to have observed ETPA have thus far been met with a large degree of caution.

One of the original ways to test for ETPA was to measure $R_{TPA}$, after a sample cell containing two-photon active chromophores, as a function of the pump laser power ($l_p$) used to produce photon-pairs *via* spontaneous parametric down conversion (SPDC). For ETPA, $R_{TPA}$ should scale linearly with $l_p$, whereas it should scale quadratically for classical two-photon absorption[15–18]. However, one-photon loss-processes such as hot-band absorption[19] or single-photon scattering[20] can also lead to the observation of a linear dependence of $R_{TPA}$ with laser power. Consequently, the above test is not sufficient to confirm ETPA. Hence, whilst additional measurements can be performed in the above scheme, *e.g.*, checking $R_{TPA}$ whilst varying the SPDC flux *after* generation [21], more sensitive and controlled schemes have recently been proposed. For example, methods based on Hong-Ou-Mandel (HOM) interferometry have been explored to detect ETPA exploiting changes in the HOM dip's visibility. However, despite several attempts[22,23], HOM-based techniques have not yet succeeded.

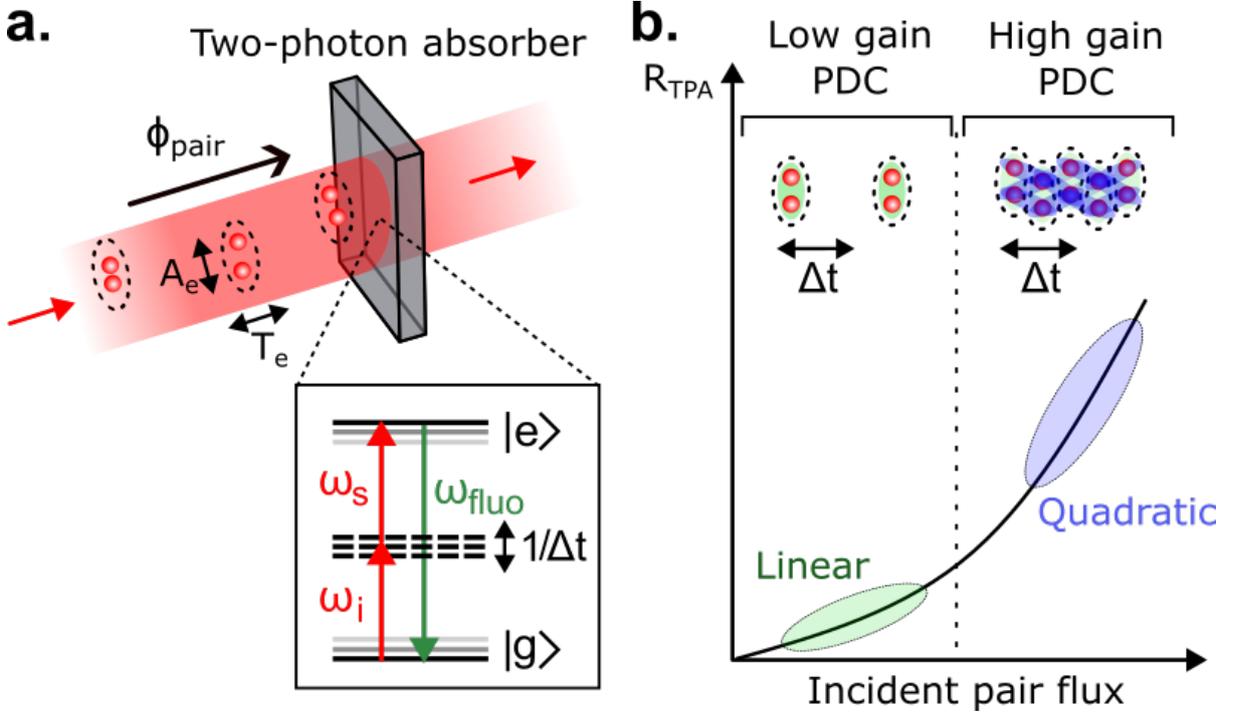

**Figure 1: Entangled two photon absorption and the response in different PDC regimes: a.** Two-photon absorber excited by entangled photon pairs. One photon excites population from the ground state $|g\rangle$ of a system to a virtual level; if a second photon is absorbed to reach the excited state $|e\rangle$, two photon-absorption occurs. The incident photon pair flux ($\phi_{pair}$), the photons energies ($\omega_i$ and $\omega_s$), and the spatial ($A_e$) and temporal ($T_e$) entanglement area are key parameters driving entangled two-photon absorption (ETPA). Ideally, $T_e$ must be much smaller than the coherence time of TPA, $\Delta t$, and $A_e$ should be minimized (*i.e.*, limited by diffraction), and both photons must have an energy that sums to the overall energy transition energy. Additionally, $\phi_{pair}$ must be low enough to ensure the average time between pairs exceeds $\Delta t$. $\omega_{fluo}$ is the frequency of the photon emitted through fluorescence. **b.** Entangled photon pairs are typically generated via spontaneous parametric down conversion (SPDC) by optical pumping of a non-linear crystal. At high pump fluences, stimulated PDC occurs and create multiple photon pairs per mode, leading to a high incident photon pair flux. Absorption can then occur from photons within the same or different pairs, not qualifying as ETPA. In this high-gain regime, the two-photon absorption rate ($R_{TPA}$) scales quadratically with pump intensity, while at low gain, it scales linearly[7–11,24,25].

The schemes described above are all based on analysis of the SPDC photons that have transmitted through a material. However, in the ideal case, one would use an intrinsic signature of the sample that is background free *e.g.*, two-photon excited fluorescence, as a probe. If the SPDC pump power is increased, a few mWs of PDC photons can be generated. This is nominally enough to induce detectable spontaneous two-photon excited emission from most chromophores that are used to test for ETPA[26]. However, in this high-gain PDC regime, photon pairs are no longer emitted in isolated time slots (with the spectrally narrow sources that are required for such experiments). Instead, the probability of multi-pair emission become significant, leading to a bright squeezed vacuum (BSV) state[27] (**Figure 1b**). The two photons involved in a TPA event can thus originate either from the same entangled pair or from different uncorrelated pairs, resulting in a pump intensity dependence that is both linear and quadratic *i.e.*, $R_{TPA} \propto \sigma_e I_p + \xi \delta_c I_p^2$, with $\xi$ a unitless parameter that can be on the order unity [28]. Both effects will give rise to an ETPA-like signature *i.e.*, two photon excited fluorescence, but only in the former case will the PL response be dominated by ETPA. Consequently, as demonstrated by Landes *et al.*[14], working in the high-gain PDC regime is not appropriate to observe ETPA. Indeed, only in the low-gain SPDC regime can entangled photon pairs arrive within a small time difference *i.e.*, have a broadband nature, whilst maintaining an energy that sums to the overall transition energy (unlike uncorrelated light), as is additionally required for ETPA.

The TPA cross-section of chromophores used for detecting ETPA have also thus-far been low. Almost all investigations have centred on molecular dyes such as Rhodamine 6g (Rh6g) or Zinc tetraphenylporphyrin (ZnTpp) with indermediate TPA cross-sections of <300 GM (GM = $10^{-50}$ cm$^4$ s photon$^{-1}$). This is despite dyes and nanomaterials existing with TPA cross-sections of >$10^5$ GM, that may more clearly exhibit the effects of ETPA[29]. Finally, it is essential to realise that $\sigma_e$ and $\delta_c$ are related by $\sigma_e \propto \frac{\delta_c}{T_e A_e}$, where $T_e$ and $A_e$ are the entanglement time and entanglement area, defined as the time and area within which photon pairs are tightly correlated[30]. Generally, minimizing both quantities is critical for boosting ETPA, but studies of the precise influence of spatial entanglement remain limited with most focusing on controlling temporal/spectral entanglement[12,23].

Here, we overcome the above discussed problems by developing and applying a new scheme to test for ETPA at room-temperature in molecular/inorganic systems (*i.e.*, not atomic gases), based on entangled photon pairs with structured spatial entanglement. We demonstrate our approach to be critically robust to one-photon losses like scattering or linear absorption. We then apply it to investigate ETPA in the low-gain SPDC regime, examining the influence of spatial entanglement and choice of chromophores on the process. We find some evidence for ETPA and characterize deviations in signals. Our results provide a steppingstone towards understanding ETPA and how it can be robustly verified and utilized.

**Main**

Our approach is based on a setup previously outlined in Cameron *et al*[31] and shown in **Figure 2a**. Spatially entangled photon pairs centred at 814 nm are generated by type-I SPDC in a β-barium borate (BBO) crystal pumped by a collimated continuous-wave laser at 407 nm. Near-degenerate down-converted photons are selected *via* spectral filters (SF) at 814 ± 2 nm. A maximum laser pump power of 100 mW is used to ensure we remain in the low-flux regime ($\leq 3 \times 10^6$ photon pairs/s). The output surface of the crystal is first Fourier-imaged onto a phase-only spatial light modulator (SLM) that is itself Fourier-imaged onto the sample plane, where the sample (or other media) are inserted. This plane is then imaged onto an electron-multiplied charge-coupled device (EMCCD) camera as shown in **Figure 2a** (for further setup details see **Supplementary Sections A** and **B**).

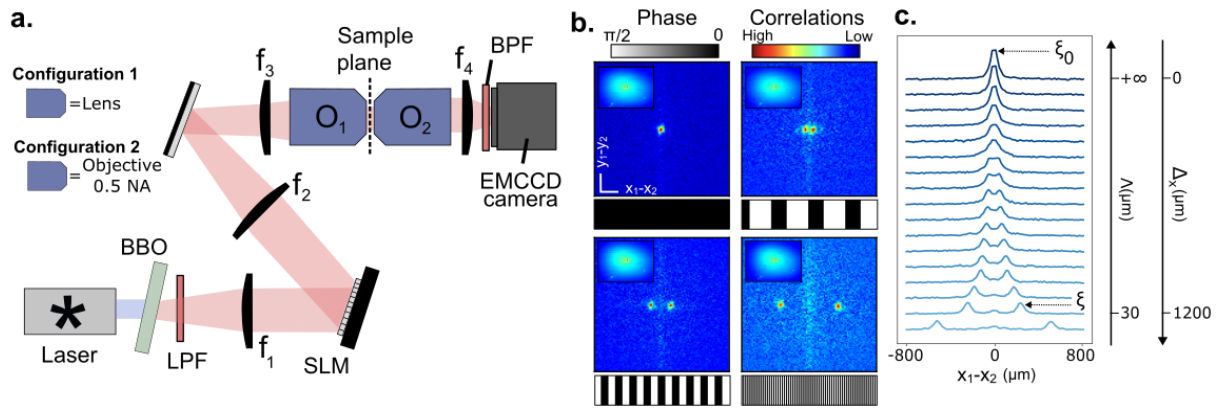

**Figure 2: Schematic of experimental setup used in this work and methodology for testing for ETPA: a.** Schematic of setup used in this work to test for ETPA. Spatially entangled photon pairs centred at 814 nm are produced *via* type-I spontaneous parametric down conversion (SPDC) using a collimated continuous-wave laser at 407 nm and a 0.5 mm thick β−Barium Borate nonlinear crystal (BBO). Pump photons are filtered out by a long-pass filter (LPF) at 650nm. A bandpass filter at 814 ± 5nm before the EMCCD filters out non-degenerate pairs. The surface of the BBO, the sample plane and the camera are in conjugate optical planes, connected by telescopes ($f_1 - f_2$, $f_3 - O_1$ and $O_2$-$f_4$). The spatial light modulator (SLM) is placed in the Fourier plane of the crystal. In 'Configuration 1', $O_1$ and $O_2$ are conventional lenses (focal length 50 mm), whereas in 'Configuration 2' microscope objectives (0.5 N.A) are used to reduce the entanglement area $A_e$ in the sample plane (see **Figure 4a**). The intensity image and the minus-coordinate projection of $G^{(2)}$ - named for clarity the *correlation image* - are both returned by the sensor after each acquisition. For further setup details see **Supplementary Sections A** and **B**. **b.** Correlation images for different SLM phase gratings (images underneath). Scale bar is 400 μm. A line of residual intensity is observed in all cases along the image vertical centre due to a constant read-out artifact ('charge smearing'; see **Supplementary Section C** for discussion). The inset shows the direct intensity image, which is the same regardless of the phase-grating. **c.** 1D line profile of the correlation image along the $x$ axis to show separation as a function of the SLM period Λ, that is inversely proportional to $\Delta_x$. $\xi_0$ and $\xi$ denote the values of the central and positive correlation peaks in the 1D line profile, respectively.

In such a setup, the photon pair beam is broad and Gaussian-shaped (area Σ) in the sample plane, as shown in the intensity images in **Figure 2b (insets).** Additionally, the photon pairs exhibit spatial correlations. These correlations are revealed by measuring the spatially resolved second-order correlation function $G^{(2)}(\boldsymbol{r_1}, \boldsymbol{r_2})$[20] and projecting it along the minus-coordinate axis (see **Supplementary Section C** and **D** for theory). This projection represents the probability of detecting two entangled photons at any pairs of pixels with positions $(x_1, y_1)$ and $(x_2, y_2)$ separated by a distance $(x_2 - x_1, y_2 - y_1)$. For clarity, we will refer to this projection as the *correlation image* throughout the manuscript. **Figure 2b** shows correlation images measured with different phase gratings programmed on the SLM. Without any phase structuring, a single peak appears at the centre, indicating strong spatial correlation between the photons: when one is detected at any position $(x_1, y_1)$ in the sample plane, its twin is most likely to be detected very close by, within a region of area $A_e$ ($A_e \ll \Sigma$). When a phase grating $\theta(\boldsymbol{r}) = \pi/4[\text{sgn}(\cos(\frac{2\pi x}{\Lambda})) + 1]$ is applied to the SLM, two peaks appear in the correlation image, separated by a distance $\Delta_x$ that is inversely proportional to the grating period Λ. In this case, if one photon is detected at any position $(x_1, y_1)$, its twin is most likely to be detected in a region centred at a distance $\pm \Delta_x/2$ from this position. Importantly, regardless of the SLM pattern, the intensity profile of the photon pair beam remains unchanged (insets). This approach thus allows for controlling the spatial arrangement of the photon pairs at each point on the sample, making them overlap ($\Delta_x = 0$) or be separated ($\Delta_x \gg \sqrt{A_e}$), without affecting the photon flux (see **Supplementary Section E** for setup calibration details).

Noting the above, we can use our setup to test for ETPA. When photons pairs are maximally overlapped ($\Delta_x = 0$), the peak in the correlation image, denoted $\xi_0$, is of maximum magnitude. If a medium is placed in the sample plane and ETPA is occurring, $\xi_0$ should be reduced as compared to when there is no medium. When the photon pairs are spatially separated at the medium plane ($\Delta_x \gg \sqrt{A_e}$), however, the probability of ETPA should be reduced to zero and hence the intensity of one of the two peaks, denoted $\xi(\Delta_x)$, should be the same with or without the medium present. If ETPA is occurring in a medium, the ratio $\xi(\Delta_x)/\xi_0$ should then be larger than in the absence of ETPA, particularly at large values of $\Delta_x$. **Figure 3a** shows experimental results demonstrating this conclusion. Here, we consider two chromophores 0.4 mM Rh6g (in methanol) or 4 monolayer thick CdSe nanoplatelets (in hexane) which have GM values of $10^1$ - $10^2$ and $10^5$ respectively at 807 nm[32,33]. Initially, both samples are placed in 1 mm pathlength borosilicate cuvettes with a total glass thickness of 2 mm. Plotting $\xi_i(\Delta x)/\xi_0$ for these samples alongside air and the non-absorbing solvent (hexane) in **Figure 3a**, interestingly shows that $\xi_i(\Delta x)/\xi_0$ follows the order CdSe>Rh6g>air/hexane, which is especially clear at large $\Delta x$ values. This suggests that we are indeed observing some ETPA occurring within these media.

To ensure that these observations do indeed arise from ETPA as opposed to setup related effects, we conducted control experiments. A critical requirement of the above scheme is that measurement of $\xi(\Delta_x)/\xi_0$ for any medium in the setup is (i) highly reproducible and that such a metric is not sensitive to (ii) linear absorption, (iii) scattering and (iv) the presence of stray light *e.g.*, background fluorescence. To test these hypotheses, we first trace out $\xi(\Delta_x)/\xi_0$ in the case when there is no medium ('air'). Throughout this manuscript measurement of $\xi_0$ and $\xi(\Delta_x)$ are repeated 4 × 1000 times with the mean $\xi_0$ and $\xi(\Delta_x)$ used to derive the $\xi(\Delta_x)/\xi_0$ ratio; the error bar is then calculated by propagation of the uncertainty on the repeat measurements of these quantities (each 1000 measurement set takes 24 h, with experiments performed over a 30-day period). The small value of the error bars (>0.01%) in **Figure 3a** ('air' curve) highlights the reproducibility and sensitivity of the presented scheme. We then compare this $\xi(\Delta_x)/\xi_0$ curve obtained for 'air' with those obtained in the presence of linear absorption and scattering. To mimic linear absorption, a linear polariser rotated by approximatively 45 degrees relative to the polarization axis of the photons is placed in the sample plane, while for scattering, we use a thin-film of parafilm (orange and light purple colour curves in **Figure 3b**, respectively). Critically, as shown in **Figure 3b**, all curves lie on-top of each other within the uncertainty of the experiment, especially at large $\delta$ values, verifying our setup is robust to one-photon loss and scattering effects. In addition, to mimic fluorescence or any stray light which could be detected in addition to the photon pairs during the acquisition, we measure $\xi(\Delta_x)/\xi_0$ with no medium whilst illuminating the camera with an 814 nm superluminescent diode (SLED); as shown by the red curve in **Figure 3b**. No statistically significant difference in $\xi(\Delta_x)/\xi_0$ is found. Finally, we also repeat experiments independently (v) reducing the flux of photon pairs illuminating the sample (by varying the pump power) and by (vi) reducing the number of pairs used in the $G^{(2)}$ measurement (by decreasing the sensor collection area), as shown in **Figure 3c** and **d**. While the signal-to-noise ratio on $\xi(\Delta_x)/\xi_0$ is reduced in these experiments, there is no quantitative difference on the behaviour of $\xi(\Delta_x)/\xi_0$ at large $\Delta x$ values, suggesting this ratio to be uniquely sensitive to properties of the medium (see **Supplementary Sections F-I** for further experimental details).

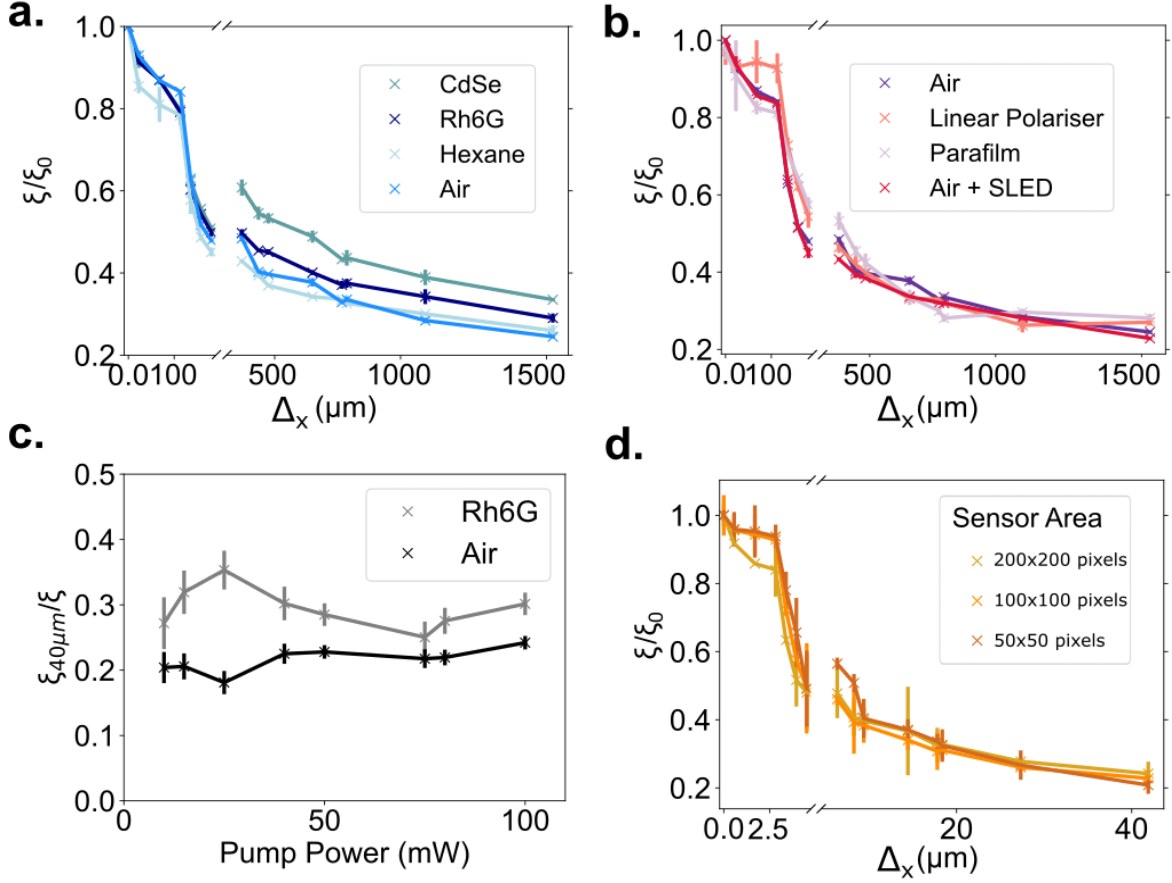

**Figure 3: Exploring ETPA with CdSe nanoplaelets and Rh6g dye checking for effects of scattering, linear absorption, pump power and collection area: a.** $\xi(\Delta_x)/\xi_0$ curve for the following media in the sample plane: 4 monolayer thick CdSe nanoplatelets (1 mg/ml concentration in hexane) in a 1 mm path length cuvette (green), 0.4 mM Rh6g (in methanol; dark blue), a 1 mm path length cuvette with hexane (light green) and 'air' (light blue). **b.** $\xi(\Delta_x)/\xi_0$ curve when in the cuvette there is 'air' (purple), a linear polariser at 45 degrees (to mimic linear absorption; orange), a parafilm layer (to mimic scattering; salmon) and 'air' together with ≈1mW of a superluminescent diode (SLED) illuminating the camera sensor (to mimic background light; red). **c.** $\xi(\Delta_x)/\xi_0$ for $\Delta_x = 40$ μm as a function of pump power for 'air' and Rh6g in the cuvette. **d.** $\xi(\Delta_x)/\xi_0$ curves for different region of interest of the EMCCD sensor with 'air' in the cuvette.

To explore the above observations further we reduce the entanglement area in the sample plane, $A_e$, from $1.72 \times 10^{-3}$ mm$^2$ in the experiments in **Figure 3** (Configuration 1) to $69.2 \times 10^{-6}$ mm$^2$, by replacing the lenses before and after the sample with 0.5 NA 20× magnification objectives (Configuration 2). Repeating measurements on Rh6g solution sandwiched and sealed between two coverslips we find there is a significant increase in $\xi(\Delta_x)/\xi_0$ when $A_e$ is decreased, as shown in **Figure 4a**, which is in-line with the theoretical prediction that the ETPA cross-section is inversely proportional to $A_e$. Together the above results indicate we are observing some ETPA.

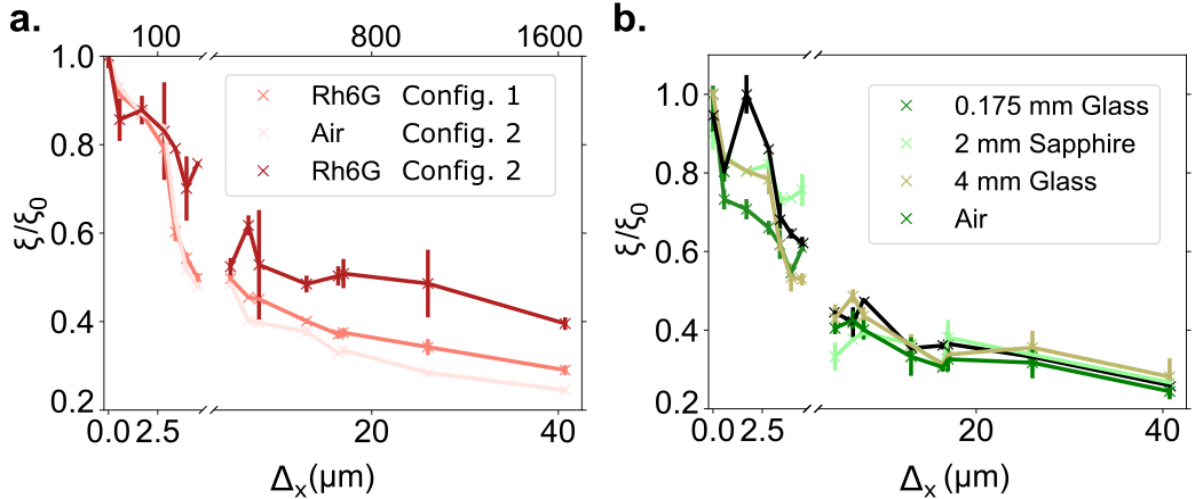

**Figure 4: Influence of entanglement area and dispersion on results: a.** $\xi(\Delta_x)/\xi_0$ curves with a cuvette of 0.4 mM Rh6g in the sample plane measured in configuration 1 (pink) and configuration 2 (green). The blue curve corresponds to a measurement with air in configuration 2. The value of $A_e$ is $1.72 \times 10^{-3}$ mm$^2$ in configuration 1 and $69.2 \times 10^{-6}$ mm$^2$ in configuration 2. The $\Delta_x$-axis scale above the graph corresponds to configuration 1, and the one below corresponds to configuration 2. **b.** $\xi(\Delta_x)/\xi_0$ curves measured for 0.175 mm thick coverslip (black), 2 mm thick sapphire plate (light blue), 4 mm thick coverslide (red) and air (green) in the sample plane. Measurements were taken in Configuration 2.

It is worth noting that one result obtained remains difficult to interpret. As shown in **Figure 3a** (pale blue and bright blue), measuring a cuvette with hexane (same result when measuring an empty cuvette) in Configuration 1, we find that the $\xi(\Delta_x)/\xi_0$ curve lies slightly above that of 'air', but well below the curve when the cuvette is filled with Rh6g/CdSe. With hexane not being known as a two-photon absorber in this wavelength range, we investigated the origin of this small – but statistically relevant – difference in more detail. For that, we firstly place media of different refractive index (2 mm sapphire plate $n$=1.79) or thickness (4 mm borosilicate glass microscope slides $n$=1.5) in the sample plane (**Figure 4b**). For the 2 mm sapphire plate we find the $\xi(\Delta_x)/\xi_0$ curve is overlapped with that of 'air', as expected. However, for the 4 mm thick glass coverslide, at large $\Delta_x$ values, $\xi(\Delta_x)/\xi_0$ is raised in a manner similar to that for the cuvette with solvent. The above observations suggest that for thick media dispersion may not influence $\xi(\Delta_x)$ and $\xi_0$ in the same way. However, this effect is relatively small, with indeed a blank coverslip (0.175 mm thickness) having the same response as 'air' in Configuration 2 (**Figure 4b**, black curve). Indeed, while the refractive index ($n$ at 800 nm) of our investigated media do follow the same order as our $\xi(\Delta_x)/\xi_0$ *i.e.*, CdSe nanoplatelets (2.45[34]) > Rh6g (1.77) > hexane (1.37), if refractive effects were playing a significant we would also expect our sapphire and Rh6g curves to be overlapped which they are not. We hence suggest refractive index differences cannot be used to explain the clear increase in $\xi(\Delta_x)/\xi_0$ at large $\Delta_x$ for CdSe/Rh6g.

**Conclusion**

In summary we have demonstrated a new measurement method based on structuring the spatial correlations between entangled photon pairs to detect ETPA at the low photon pair fluxes. We have highlighted that our scheme is robust to common artefacts that typically plague ETPA measurements including the presence of stray light such as background fluorescence, one-photon loss processes (linear or hot absorption) and scattering. We find that with our scheme an effect resembling ETPA can be observed for Rh6g and CdSe chromophores, irrespective of the transverse correlation width of photon pairs, SPDC flux or detection parameters.

The confidence in our results relies on the extreme robustness and sensitivity of our measurement method, including the choice of the metric $\xi(\Delta_x)/\xi_0$. Operating the source at low gain and maintaining a constant total photon flux regardless of the spatial correlation structure of photon pairs, the probability of two photons from different pairs being absorbed by the sample is not only very low but also not expected to vary with $\Delta_x$. In addition, our $G^{(2)}$ measurement method involves subtraction of accidental coincidences, so the measured correlation peak height is proportional only to the rate of genuine entangled photon pairs. The differences observed between the curves thus cannot be attributed to the absorption of two photons from different pairs, but rather to the disappearance of entangled photon pairs, suggesting ETPA.

Our work represents a new stepping stone towards verifying ETPA in condensed phase chemical systems, at room-temperature, more widely. While our results are promising, given the conflicting results and challenges that have been reported when detecting ETPA, we proceed with caution. On the setup side there remains further work *e.g.*, to better characterise dispersion effects by switching from lens-based setups to ones with only mirrors. which could make the above methods quantitative to accurately determine ETPA cross-sections. Also, the effect of the phase between photons in the pair[35] (see **Supplementary Section H**), and testing chromophores with higher TPA cross-sections would be important to investigate. Similarly, it will be important to explore sum-frequency generation with our scheme and incorporate additional measurements *e.g.*, of the joint spectral amplitude, to determine which exact parameters are most sensitive to ETPA effects.

It remains exciting to see if biological media stained with two-photon absorbers can be imaged with photon pairs at lower photon fluxes than with classical light (for an equivalent signal-to-noise ratio)[36]. Beyond two-photon absorbers, the scheme we have detailed could also be used to explore entangled light-matter interactions in other systems[37] *e.g.*, polaritonic microcavities or plasmonic materials or for lithographic patterning[38]. However, increasing the flux of entangled photon pairs while continuing to operate in a photon-sparse *(i.e.*, $\leq 1$ pair per temporal mode) regime will be critical for exploring any of the above problems. In this respect, the use brighter and purer sources of photon pairs not relying on SPDC, such as quantum dots[39], could play an important role in this research efforts.

Supplementary material: **Towards robust detection of entangled two-photon absorption**

A. Additional details in the experimental setup

**Figure S1 a** and **b** show full details of the experimental configurations 1 and 2, respectively. The BBO crystal, measuring $0.5 \times 5 \times 5$ mm, is cut for type I SPDC at 405 nm with a half-opening angle of 3 degrees (Newlight Photonics). To achieve near-collinear phase matching at the output (where the ring collapses into a disk), it is slightly rotated around the horizontal axis. The pump source is a continuous-wave laser at 407 nm (Coherent OBIS-LX) with an output power of approximately 100 mW and a beam diameter of $0.8 \pm 0.1$ mm. After the crystal, a 650 nm-cut-off long-pass filter is used to block the pump photons, along with a band-pass filter centred at $814 \pm 2$ nm. The camera used is an EMCCD (Andor Ixon Ultra 897) operating at –60 °C, with a horizontal pixel shift readout rate of 17 MHz, a vertical pixel shift of 0.3 µs, a vertical clock amplitude voltage of +4V above the factory setting, and an amplification gain set to 1000. The full camera sensor consists of $512 \times 512$ pixels with a 16 µm pixel pitch and nearly unity fill factor. The exposure time is set to 2 ms, and the camera captures approximately 100 frames per second (fps) within a $150 \times 150$ pixel region of interest. This region of interest size was used in all experiments, except in **Figure 3d** of the manuscript where it was intentionally varied. The lens $f_1$ performing the Fourier transform between the crystal surface and the SLM in Figure 2a of the manuscript is depicted with one lens for simplicity, but actually consists of three lenses with focal lengths of $f_a = 40, f_b = 150$ and $f_c = 100$ mm. The two-lens imaging system $f_2 - f_3$ in **Figure 2a** of the manuscript is consists of four lenses with focal lengths of $f_d = 50, f_e = 75, f_f = 100$ and $f_g = 100$ mm. The lens $f_4$ in Figure 2a of the manuscript is consists of one lens with focal length of $f_h = 100$ mm additionally. In Configuration 1, $O_1$ and $O_2$ are converging lenses with a focal length of 50 mm. They are positioned so that they are at the focal distance from the sample plane. In configuration 2, they are replaced by 0.5 NA microscope objectives (Nikon, Plan Fluorite), positioned so that both are at the working distance from the sample plane. The magnifications between the sample plane and the camera are $M_1 = 2$ and $M_2 = 10$ in Configuration 1 and Configuration 2, respectively.

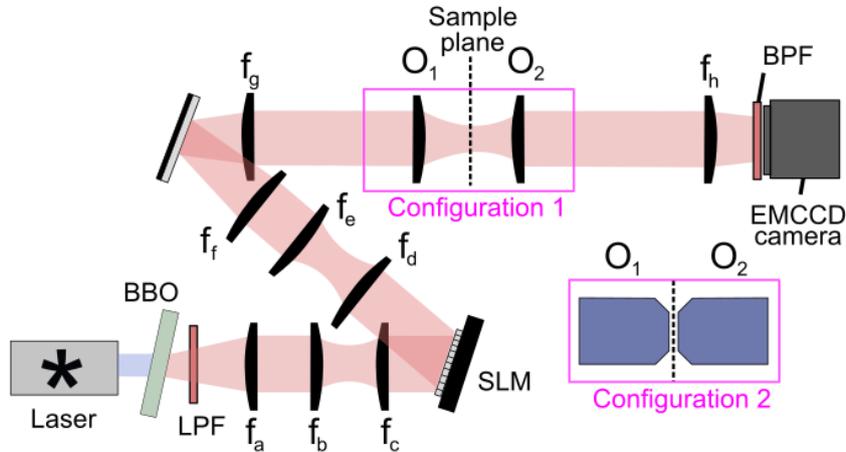

**Figure S1.** Detailed experimental setups in configuration 1 and 2. In Configuration 1, $O_1$ and $O_2$ are converging lenses with a focal length of 50 mm. In configuration 2, they are replaced by 0.5 NA microscope objectives (Nikon, Plan Fluorite). The magnifications between the sample plane and the camera are $M_1 = 2$ and $M_2 = 10$ in Configuration 1 and Configuration 2, respectively.

## B. Properties of the spatially entangled photon pair source

In our experiment, a thin-BBO crystal is illuminated using collimated 0.8 mm-diameter 407 nm CW laser with a power of 100 mW. In such case, SPDC is occurring in the low gain regime, which means that the production of more than two photons per temporal mode at the output can be neglected. In addition, the bandpass filter selects only the degenerate photons at 814 nm, which allows us to neglect spatio-spectral coupling and consider degenerate and quasi-monochromatic photons. Assuming a pure state, the two-photon states thus reads:

$$|\Psi\rangle \approx |0\rangle + \eta \iint \phi(r_1, r_2) \, a^+_{r_1} a^+_{r_2} |0\rangle \quad (A1)$$

where $\phi(r_1, r_2)$ is the spatial two-photon wavefunction. Using the double-Gaussian approximation[40,41], this function can be expressed in the sample plane, which is conjugate with the crystal surface, as follows:

$$\phi(r_1, r_2) = A \exp\left(-\frac{|r_1 - r_2|^2}{4 A_e}\right) \exp\left(-\frac{|r_1 + r_2|^2}{4(2\Sigma - A_e)}\right) \quad (A2)$$

where $r_1 = (x_1, y_1)$ and $r_2 = (x_2, y_2)$ are the transverse positions of the photons, $A_e$ is the entanglement area and $\Sigma$ is the overall beam area in the sample plane (note that $A_e$ and $\Sigma$ are related to the correlation widths in position and momentum, variables that are commonly used in other studies involving spatially entangled photon pairs [41]). In practice, the values of $A_e$ and $\Sigma$ depend on the crystal parameters, pump beam diameter, and magnification between the crystal surface and the sample plane. In our work, we estimate their values experimentally using Gaussian fits. **Figure S2** shows the intensity and correlation images used to estimate the values of $A_e$ and $\Sigma$ in both configurations, and **Table 1** shows their measured values.

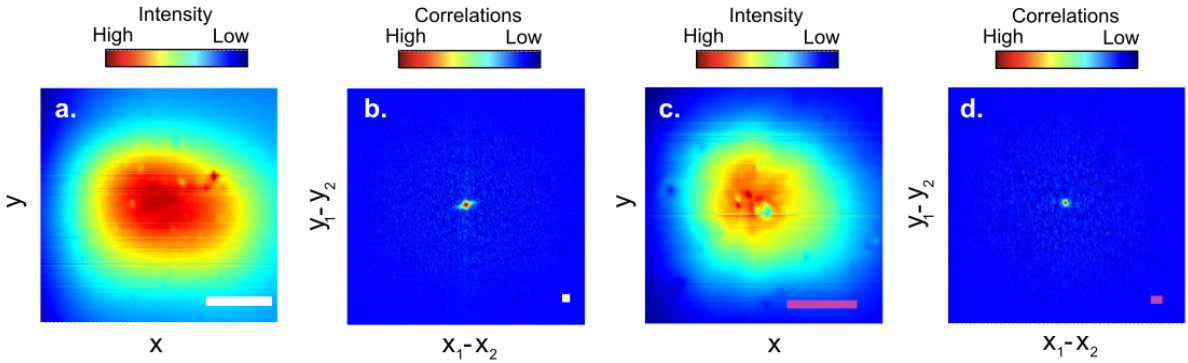

**Figure S2:** **(a)** Intensity image and **(b)** correlation image in Configuration 1. **(c)** Intensity image and **(d)** correlation image in Configuration 2. The areas $\Sigma$ and $A_e$ are first measured on the camera by fitting the images with Gaussian functions of the form $\exp\left(-\frac{x^2+y^2}{2 A_e}\right)$ and $\exp\left(-\frac{x^2+y^2}{2 \Sigma}\right)$. Their radii are then converted to values in the sample plane by dividing by the magnifications $M_1 = 2$ and $M_2 = 10$. Scale

bars in (a) and (b) are 1.6 mm and 48 μm respectively. In (c) and (d) the scale bars are 1.1 mm and pink scale bar are 32 μm, respectively, expressed in the camera plane.

| Configuration | $A_e$ (mm²) | $\Sigma$ (mm²) |
|---|---|---|
| 1 | $1.72 \times 10^{-3}$ | 1.92 |
| 2 | $69.2 \times 10^{-6}$ | 0.0432 |

**Table 1:** Measured values of $A_e$ and $\Sigma$ in configuration 1 and 2.

## C. Measuring spatially resolved $G^{(2)}$ with an EMCCD camera.

*a. Measurement process*

Given a set of M frames acquired by the EMCCD camera, one can compute (1) the intensity distribution $I(i,j)$ at pixel $(i,j)$ and (2) the intensity correlations distribution $G^{(2)}(i,j,k,l)$ between pixels $(i,j)$ and $(k,l)$:

(1) The intensity distribution is computed by averaging the intensity over all the frames: $I(i,j) = \frac{1}{M}\sum_{m=1}^{M} I_m(i,j)$ where $I_m(i,j)$ is the intensity value measured at pixel $(i,j)$ of the $m^{th}$ frame.
(2) The second-order intensity correlation distribution between pixels $(i,j)$ and $(k,l)$ is computed by multiplying the values from the same frame of the two pixels and subtracting the product of the value of the two pixels from consecutive frames. This is then averaged over all the frames and reads:

$$G^{(2)}(i,j,k,l) = \frac{1}{M}\sum_{i=1}^{M} I_m(i,j)I_m(k,l) - \frac{1}{M-1}\sum_{i=1}^{M-1} I_m(i,j)I_{m+1}(k,l) \qquad (A3)$$

The first term in equation A3 originates from detections of both genuine photon coincidence (two photons from the same entangled pair) and accidental coincidence (two photons from two different pairs, or noise). Because there is zero probability for two photons from the same entangled pair to be detected in two successive images, the second term originates only from accidental coincidences. In our experiment, since the exposure times used (~ms) are much larger than the inverse of the photon-pair rate (~ $10^6$ pairs per second), the accidentals are dominating in the first term. Subtracting it the second term – which effectively corresponds to calculating a covariance – allows to remove the accidentals and conserve only the genuine correlations. When the light detected is coming from an SPDC source operated in the low-gain regime, it was shown in Ref.[42] that the result of this subtraction is exactly equal to the spatially-resolved second order correlation function $G^{(2)}(i,j,k,l)$ of the photon pairs.

In practice, the measured $G^{(2)}(i,j,k,l)$ takes the form a 4-dimensional matrix containing $(N_x \times N_y)^2$ elements, where $N_x \times N_y$ correspond to the region of interest of the camera. It corresponds to the probability of detecting jointly two photons from the same pair at pixels $(i,j)$ and $(k,l)$. It is a discrete version of $G^{(2)}(\boldsymbol{r_1},\boldsymbol{r_2})$ that is by definition the absolute square of the spatial two-photon wavefunction of the two-photon state, where $\boldsymbol{r_1} = (x_1,y_1)$ and $\boldsymbol{r_2} = (x_2,y_2)$ are continuous transverse positions. In order to take advantage of the specific symmetries of $G^{(2)}$ related to the experimental configuration used in our work, we visualize it in practice by projecting it along the minus-coordinates axis, defined as:

$$\Gamma^-(\boldsymbol{\delta r^-}) = \iint G^{(2)}(\boldsymbol{\delta r^-} - \boldsymbol{r}, \boldsymbol{r})\mathrm{d}\boldsymbol{r} \qquad (A4)$$

with $\boldsymbol{\delta r^-} = \boldsymbol{r_1} - \boldsymbol{r_2}$. Using the measured $G^{(2)}$ in the discrete-variable formalism, this becomes:

$$\Gamma^-(i^-, j^-) = \sum_{i=1}^{N_x} \sum_{j=1}^{N_y} G^{(2)}(i^- - i, j^- - j, i, j). \quad (A5)$$

*b. Artefacts removal*

Using an EMCCD camera, not all correlation values of $G^{(2)}$ can be measured using Equation A3, in particular those (a) at the same pixel *i.e.*, $G^{(2)}(i, j, i, j)$ and (b) those between vertical neighbouring pixels *i.e.*, $G^{(2)}(i, j, i, j \pm 1)$:

(a) Since the EMCCD camera cannot resolve the number of photons incident on a single pixel, equation A3 is not valid for $(i, j) = (k, l)$. In practice, this issue is solved by interpolating all the values in the diagonal $G^{(2)}(i, j, i, j)$ using neighbouring correlation values on the same row: $[G^{(2)}(i, j, i + 1, j) + G^{(2)}(i, j, i - 1, j)]/2 \to G^{(2)}(i, j, i, j)$.

(b) Because of the so-called smearing effect[43], artefacts due to crosstalk appear in $G^{(2)}$ between pixels located on the same column. In practice, these values are interpolated using neighbouring correlation values on the same row: $[G^{(2)}(i, j, i - 1, j \pm 1) + G^{(2)}(i, j, i + 1, j \pm 1)]/2 \to G^{(2)}(i, j, i, j \pm 1)$.

More details on the $G^{(2)}$ measurement process and artifacts are found in Refs.[42,43]

**D. Theory of structuring spatially entangled photon pairs with an SLM**

When a phase mask $\theta(r')$ is displayed on the SLM, it is shown in Cameron *et al.*[31] that $G^{(2)}(r_1, r_2)$ of incident photons in the sample plane becomes

$$G^{(2)}(r_1, r_2) = \left| g * \mathcal{F}[e^{i\psi(r')}] \right|^2 (r_1 - r_2) \quad (A6)$$

where, $\mathcal{F}[...]$ is the two-dimensional Fourier transform operator, $g(r) = e^{-\frac{r^2}{4A_e}}$, $\psi(r') = \theta(r') + \theta(-r')$, $r$ and $r'$ being the spatial coordinates in the sample and SLM plane, respectively. Hence, the SLM mask and its spatially inverted version will shape the photon-pair correlations ($G^{(2)}$). If we consequently consider a $[0, \pi/2]$ phase grating on the SLM of the form $\theta(r') = \pi/4[\text{sgn}(\cos(\frac{2\pi x'}{\Lambda})) + 1]$, where sgn is the sign function and $\Lambda$ is a controllable grating period, we observe a splitting of the central correlation peak into two symmetric peaks separated by a distance $\Delta_x \sim 1/\Lambda$, as shown in **Figure 2b** of the manuscript. The phase grating thus shapes the photon pairs so that they become spatially separated in the sample plane. By controlling its frequency, we can monotonically go from a regime where pairs are tightly overlapped in the medium plane to forcing pairs to be separated from one another by a chosen distance $\Delta_x$. Throughout this process, we note that the area of the peaks remains constant and equal to $A_e$, akin to the intensity, which exhibits minimal variation and maintains its gaussian shape of area $\Sigma$. However, the height of the peaks, $\xi$, slightly decreases (see **Figure 2c** of the manuscript), which is mainly due to the imperfect diffraction efficiency of the SLM.

*a. Theoretical derivation of equation (A6)*

This section provides a derivation of equation (*A6*) that is noted above. For that, we consider the propagation of the two-photon wavefunction from the SLM plane to the sample plane:

$$G^{(2)}(r'_1, r'_2) = \left| \iint \psi_{SLM}(r_1, r_2) \, e^{i\theta(r_1)} e^{i\theta(r_2)} h(r'_1, r_1) h(r'_2, r_2) dr_1 dr_2 \right|^2 \quad (A7)$$

where $G^{(2)}(r'_1, r'_2)$ is the second-order correlation function in the sample plane, $\psi_{SLM}(r_1, r_2)$ is the two-photon wavefunction in the SLM plane, $r'_1$ and $r'_2$ ($r_1$ and $r_2$) are the transverse positions of the photons in the sample plane (SLM plane), $\theta$ is the phase mask programmed on the SLM and $h(r', r) = e^{i\frac{r'r}{\lambda f}}$ is the coherent point spread function associated with the optical Fourier transform performed by the lenses and/or microscope objective between the SLM and sample plane ($\lambda$ is the wavelength of the photons and $f$ is an effective focal length). In addition, under our experimental conditions, we assume that the photon pairs are near-perfectly anti-correlated in the SLM plane *i.e.*, $\psi_{SLM}(r_1, r_2) \approx \psi_0(r_1 - r_2)\delta(r_1 + r_2)$, where $\delta$ is the Dirac delta function and $\psi_0$ is the envelope of the two-photon wavefunction in the SLM plane. This so-called thin crystal approximation[41,44] is valid in our work because we use a collimated pump with a diameter larger than the crystal thickness. It remains valid as long as the spatial frequencies of the phase patterns programmed on the SLM are smaller than the inverse of the correlation width in the SLM plane. In our experiment, we measured a correlation width of 0.34 mm in the SLM plane, much smaller than the minimum SLM grating period of 1.3 mm, confirming the validity of the approximation (see **Figure S3**). Furthermore, since the crystal is slightly tilted to produce a disk rather than a ring in the SLM plane, we can use the double Gaussian approximation and model the envelope $\psi_0$ in the following Gaussian form: $\psi_0(r) = e^{-\frac{|r|^2}{4\Sigma_{SLM}}}$, where $\Sigma_{SLM}$ is the area of the beam in the SLM plane. $\Sigma_{SLM}$ depends mainly on the crystal thickness and the lenses between the crystal surface and the SLM performing the Fourier transform. Under this hypothesis, the previous equation simplifies to:

$$G^{(2)}(r'_1, r'_2) = \left| \iint e^{-\frac{|r_2 - r_1|^2}{4\Delta_{SLM}}} \delta(r_1 + r_2) e^{i\theta(r_1)} e^{i\theta(r_2)} e^{i\frac{r'_1 r_1}{\lambda f}} e^{i\frac{r'_2 r_2}{\lambda f}} dr_1 dr_2 \right|^2$$

$$= \left| \int e^{-\frac{|r|^2}{4\Sigma_{SLM}}} e^{i(\theta(r) + \theta(-r))} e^{i\frac{(r'_1 - r'_2)r}{\lambda f}} dr \right|^2$$

$$= \left| \mathcal{F}\left[e^{-\frac{|r|^2}{4\Sigma_{SLM}}}\right] * \mathcal{F}[e^{i\psi(r)}] \right|^2 (r'_1 - r'_2)$$

$$= \left| g * \mathcal{F}[e^{i\psi(r)}] \right|^2 (r'_1 - r'_2) \qquad (A7)$$

where $g(r) = e^{-\frac{|r|^2}{4A_e}}$, with $A_e$ is the correlation width in the sample plane, and $\psi(r) = \theta(r) + \theta(-r)$.

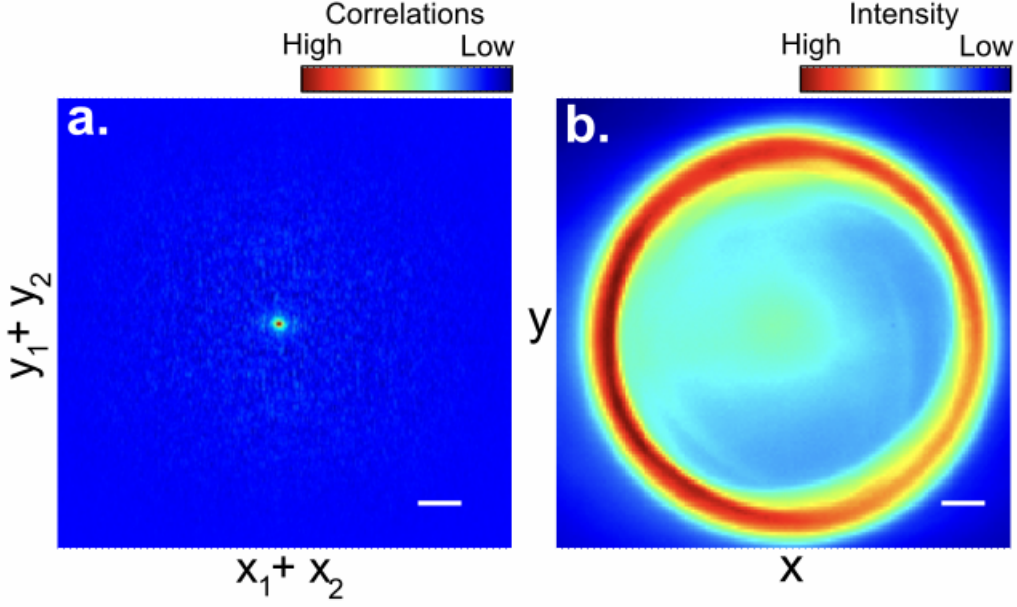

**Figure S3: a-b.** Correlation **(a)** and intensity (b) images acquired by Fourier imaging the crystal onto the camera. Scale bars are 1.5 mm in both images. The correlation image corresponds to the sum-coordinate projection of $G^{(2)}$. From these images and lenses used in the measurement configuration, we can estimate a correlation width in the SLM plane of approximately 0.34mm.

   b. *Phase grating modulation*

Considering the phase $\theta(\boldsymbol{r}) = \pi/4 \left[ sgn\left(\cos\left(\frac{2\pi x}{\Lambda}\right)\right) + 1 \right]$, the Fourier transform in the previous formula becomes:

$$\mathcal{F}[e^{\pi/4i\left[sgn(\cos(\frac{2\pi x}{\Lambda}))+sgn(\cos(-\frac{2\pi x}{\Lambda}))+2\right]}](\boldsymbol{r}') = i\mathcal{F}[e^{\pi/2i\, sgn(\cos(\frac{2\pi x}{\Lambda}))}](\boldsymbol{r}')$$

$$= -\sum_{n=-\infty}^{+\infty} 2\, c_{2n+1}\, \delta\left(x' - (2n+1)\frac{\lambda f}{\Lambda}\right) \quad (A8)$$

where $c_n = \frac{1}{2} sinc\left(\frac{n}{2}\right)$. The second order correlation function in the sample plane can thus be written:

$$G^{(2)}(\boldsymbol{r'}_1, \boldsymbol{r'}_2) = \left| \sum_{n=-\infty}^{+\infty} c_{2n+1}\, e^{-\frac{\left|x'_1 - x'_2 - (2n+1)\frac{\lambda f}{\Lambda}\right|^2}{4 A_e}} e^{-\frac{|y'_1 - y'_2|^2}{4 A_e}} \right|^2 \quad (A9)$$

This result is in accordance with the experimental minus-coordinate projection of $G^{(2)}$ shown in **Figure 2b** of the manuscript.

**E. Experimental calibration of the correlation shaping process**

As shown by the previous equation, $G^{(2)}(\boldsymbol{r'}_1, \boldsymbol{r'}_2)$ has no zero-order term ($c_0 = 0$), no even order terms, and is dominated by the first order terms ($c_1 = c_{-1} \approx 0.32$). This is observed experimentally in **Figure 2b** of the manuscript. To obtain this result, however, one must carefully calibrate beforehand the lateral shift $\alpha$ in the programmed phase mask i.e. $\theta(\mathbf{r}) = \pi/4 \left[ sgn(\cos\left(\frac{2\pi x}{\Lambda} + \alpha\right)) + 1 \right]$, where $\alpha$ is the lateral shift, to make it precisely equal to zero. **Figure S3** shows minus-coordinate projections for a fixed $\Lambda = 300\mu m$ and different values of $\alpha$. In practice, we find the best value of $\alpha$ by testing many shifts and selecting the one that minimize the zero-order peak.

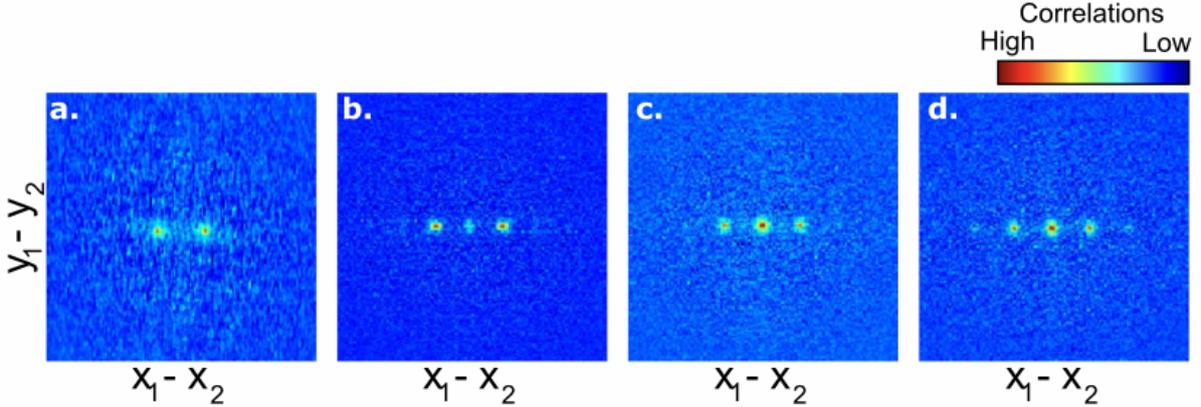

**Figure S4:** Minus-coordinate projections (with no medium) for a fixed $\Lambda = 300$ μm and different values of $\alpha$ (the lateral shift applied to the SLM phase grating). Panel (a) $\alpha = 0.34$; panel (b); $\alpha = 0.38$ panel (c) $\alpha = 0.41$; panel (d) $\alpha = 0.44$. This approach allows us to calibrate the SLM and minimise the zero (or other) order peak(s).

### F. Details on the measurement of $\xi$

$\xi$ is defined as the height of the correlation peak visible on the correlation image. This peak, by definition, represents the sum of all coincidences of entangled photon pairs incident on the same pixel, detected within the camera's region of interest, making $\xi$ directly proportional to the total number of detected entangled photon pairs. It is a crucial metric in our study. Specifically, $\xi_0$ represents the peak value when no phase mask is applied to the SLM, while $\xi(\Delta_x)$ is the value of the shifted peak on the positive side. Note that, given that the correlation image is inherently centrosymmetric, any of the peaks can be used to define $\xi(\Delta_x)$.

In all our measurements, we evaluate the ratio $\xi_0/\xi(\Delta_x)$ for various values of $\Delta x$. The resulting curve consistently decreases. For 'air' (depicted as the light blue curve in **Figure 3a** of the manuscript), the ratio decreases from 1 to approximately 0.25, which is close to the theoretically predicted value of $c_1 \approx 0.32$. The slight discrepancy can be attributed to the fact that, in practice, the SLM shaping is not perfect, as some portion of the light remains uncontrolled and ends up in the zero order, thereby reducing the ratio. Finally, we note that other metrics such as the area of the correlation peaks in the correlation result in the exact same trends in $\xi_0/\xi(\Delta_x)$.

### G. Metric robustness

*a. Insensitivity to noise and stray light*

The insensitivity to noise and stray light is related to the method of detecting photon pairs through their spatial correlations as detailed in Section C of the supplementary document. The signal measured, *i.e.*, the peak in the correlation images, is formed solely by the spatial correlations between genuine entangled photon pairs. Indeed, all accidental coincidences are removed in the subtraction of equation *A3*. Thus, noise events or photons from stray and ambient light, which appear as accidental coincidences, do not contribute to the signal. They only increase the noise in the correlation image, which practically reduces the signal-to-noise ratio. This effect was previously observed and quantified in Ref.[45] In our work, this is clearly visible in **Figure 3b** of the manuscript, where the presence of stray light from the SLED does not alter the shape of the measured curve. Note that this remains true provided that the stray light falling on the camera sensor is of sufficiently low intensity to avoid saturating it, which is a reasonable assumption in our study.

b. *Insensitivity to linear losses*

This insensitivity to linear losses is related to the calculation of a ratio between two quantities, $\xi_0$ and $\xi(\Delta_x)$, which are affected in the same way by linear losses (*i.e.*, one-photon losses). Indeed, the probability of losing a photon through this type of absorption is independent of whether the photon is spatially correlated with another photon or not. Thus, $\xi_0$ and $\xi(\Delta_x)$ which are two quantities proportional to the number of entangled pairs detected by the camera, are attenuated in the same proportion in the presence of single-photon losses, allowing the ratio $\xi_0/\xi(\Delta_x)$ to remain constant. This insensitivity is confirmed by measurement shown in **Figure 3b** of the manuscript.

c. *Insensitivity to scattering*

This insensitivity to linear losses is related to the calculation of a ratio between two quantities, $\xi_0$ and $\xi_i(\Delta x)$, which are affected in the same way by scattering. The presence of scattering causes a dispersion of the correlation peak in the correlation image. Indeed, photon pairs effectively lose their strong spatial correlations. In cases of significant scattering (*e.g.*, multiple scattering), the peak spreads completely across the entire correlation image and can become indistinguishable from noise, as observed in Ref.[46]. In cases of weak scattering - which could be the case in our experiment - the peak slightly spread and lose intensity. Ultimately, regardless of the strength of the scattering, it is an effect that acts on each photon individually, and therefore has the same effect on $\xi_0$ and $\xi(\Delta x)$. The ratio $\xi_0/\xi(\Delta x)$ thus remains constant. This insensitivity is confirmed by measurement shown in **Figure 3b** of the manuscript.

**H. Variation with the phase**

It has been suggested by Bi *et al.*[35] that ETPA can be enhanced when there is a large phase difference between photons in a pair. To test this, we ramp the phase on one half of the SLM, where the beam impinges, between 0 and $\pi/2$ while keeping the other half of the SLM at fixed phase (of 0). Since the photons in a pair are strongly anti-correlated in momentum (i.e. $\vec{k}_{\perp_s} \approx -\vec{k}_{\perp_i}$), they arrive on the SL in opposite halves of the beam, so this allows us to control their relative phase. Measuring $\xi_0$ (in Setup 1) for Rh6g, air and 1 mm thickness borosilicate glass (**Figure S5**) we find no difference in the response regardless of the phase difference between pairs. This suggest that further work is required *e.g.*, tuning the spectral bandwidth of the photon pairs to understand the role of photon phase in ETPA.

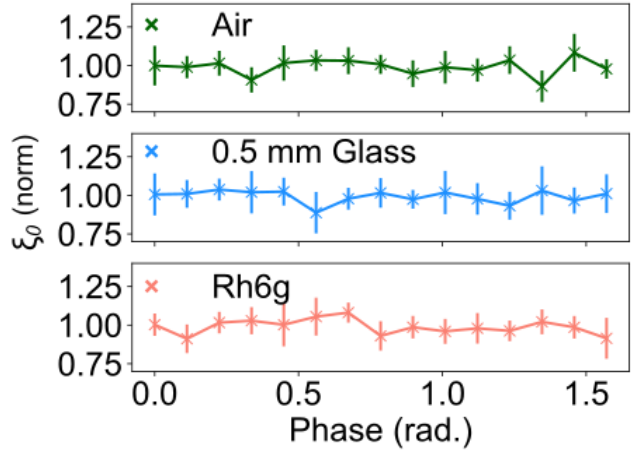

**Figure S5:** $\xi_0$ as a function of the SLM phase on one half of the beam; the other half is kept at fixed phase (0). Measurements are taken for air (green), 0.5 mm thick glass coverslide (green) and cuvette with 0.4 mM Rh6g (salmon) in the NF medium plane. $\xi_0$ is normalised to its value when the phase difference is zero.

## I. Errorbars

In the main text we display effectively a $\langle\xi\rangle$ for each $(\Delta x)$. In each case this involves measuring $\xi_i$ 1000 times and then repeating this on 4 occasions. The error bar is then determined from the standard error on all such $\xi_i$ values measured and propagated according to $\Delta(\xi/\xi_0) = (\xi/\xi_0) \times [(\Delta(\xi_0)/\xi_0) + (\Delta(\xi)/\xi)]$, where $\Delta$ is the standard error on the given quantity in brackets. The measurement of $\xi_i$ is exceptionally reproducible and effectively a property of the measurement system, this gives us confidence in comparing curves of $\xi/\xi_0$ effectively even when they lie close to one another.

## J. Variables

$R_{TPA}$: two-photon absorption rate
$\sigma_e$: entangled two-photon absorption cross section
$\phi_{pair}$: photon pairs flux
$\delta_C$: classical two-photon absorption
$l_p$ : pump power
$\Sigma$: photon pair beam diameter in the sample plane
$\Lambda$ : Grating period in the SLM plane
$\Delta_x$ : Distance between the two correlation peaks in the sample plane
$T_e$ : entanglement time of entangled photon pairs in the sample plane
$A_e$: entanglement area of entangled photon pairs in the sample plane
$\xi_0$ : value of the correlation peak for $\Delta_x = 0$.
$\xi(\Delta_x)$: value of the positive correlation peak for $\Delta_x \neq 0$.

## I. Chemistry

*a. CdSe Nanoplatelet Synthesis*
All chemicals were used as received: Octadecene (ODE) (Sigma-Aldrich, 90%), cadmium acetate dihydrate (Cd(Ac)$_2$.2H$_2$O) (Sigma-Aldrich, 98%), cadmium oxide (CdO) (Strem 99.99%), myristic acid (Sigma-Aldrich, 99%), selenium powder (Se) (Strem Chemicals 99.99%), oleic acid (OA) (Sigma-Aldrich, 90%), oleylamine (OLA, Acros, 80−90%), n-hexane (VWR, 99%), methanol (MeOH) (VWR, 99.8%) and ethanol absolute (VWR, 99.8%) are used.

The cadmium acetate dihydrate was used as purchased and crushed and then dried under vacuum at 70°C overnight.

The Cd(myristate)$_2$ was synthesized by loading 2.56 g (20 mmol) of CdO and 11 g (50 mmol) of myristic acid into a 50 mL three-neck flask. The mixture in the flask was heated to 80°C and degassed for 30 min. Under an argon flow, the solution was kept heating at 200°C until becoming colorless. During the cooling step, 30 mL of MeOH was added at 60°C in order to solubilize the excess myristic acid. The obtained white solid was precipitated 5 times with MeOH using a centrifuge tube. The final product was dried at 70°C under vacuum overnight.

4 monolayer ML 20 nm × 8 nm zinc-blende CdSe NPLs were synthesized using a slightly modified version of the procedure from Ithurria *et al.* In a 25 mL three-neck flask, 170 mg of Cd(myristate)$_2$, 12 mg of Se powder and 15 mL of ODE were loaded and degassed under vacuum for 30 min at room temperature. The mixture was then heated to 240°C under an argon flow. At around 200°C, 90 mg of dried Cd(Ac)$_2$ was quickly added to the intense-colored orange solution. The reaction was stopped when the 512 nm absorption was stabilized (around 20 min after the addition of Cd(Ac)$_2$). During the cooling step, 150 μL of OA was injected into the solution at 150°C. After cooling, the NPLs were purified twice with 15 mL of hexane and 15 mL of ethanol by centrifuging at 6000 rpm for 5 min. the platelets presented a band-edge at 512 nm.

b. Rhodmine 6G

Rhodamine 6G was used as purchased (Sigma-Aldrich) and prepared to 5 mM concentration in ethanol (Sigma-Aldrich, 99.9%).

c. Sample preparation

For experiments using 'Setup 1' detailed in Figure 2a of the main text the Rhodamine 6G, CdSe nanoplatelets or hexane were placed in a 1 mm path length quartz cuvette. For experiments with 'Setup 2' solutions of Rhodamine 6G, CdSe nanoplatelets or hexane were placed in home-made cuvettes made by sandwiching two-pieces of 0.17 mm thick coverglass together with an approximately 100 μm thick spacer. Care was taken to ensure solvent/sample did not evaporate from the cuvettes significantly throughout the course of experiments and fresh cuvettes were typically made between repeat experiments to ensure sample degradation did not influence the results[47].

**Bibliography**

1. Pawlicki, M., Collins, H. A., Denning, R. G. & Anderson, H. L. Two-Photon Absorption and the Design of Two-Photon Dyes. *Angewandte Chemie International Edition* **48**, 3244–3266 (2009).
2. Akemann, W. *et al.* Two-photon voltage imaging using a genetically encoded voltage indicator. *Sci Rep* **3** 2231 (2013)
3. Wu, E.-S., Strickler, J. H., Harrell, W. R. & Webb, W. W. Two-photon lithography for microelectronic application. SPIE Procceedings **1674**, 776–782 (1992).
4. Qin, Y., Schnedermann, C., Tasior, M., Gryko, D. T. & Nocera, D. G. Direct Observation of Different One- And Two-Photon Fluorescent States in a Pyrrolo[3,2-b]pyrrole Fluorophore. *Journal of Physical Chemistry Letters* **11**, 4866–4872 (2020).
5. Schlawin, F., Dorfman, K. E. & Mukamel, S. Entangled Two-Photon Absorption Spectroscopy. *Acc Chem Res* **51**, 2207–2214 (2018).
6. Lerch, S. & Stefanov, A. Experimental requirements for entangled two-photon spectroscopy. *Journal of Chemical Physics* **155**, 1ENG (2021).
7. Boitier, F. *et al.* Photon extrabunching in ultrabright twin beams measured by two-photon counting in a semiconductor. *Nature Communications 2011* **2**, 1–6 (2011).
8. Dayan, B., Pe'er, A., Friesem, A. A. & Silberberg, Y. Nonlinear interactions with an ultrahigh flux of broadband entangled photons. *Phys Rev Lett* **94**, 043602 (2005).
9. Dayan, B., Pe'er, A., Friesem, A. A. & Silberberg, Y. Two photon absorption and coherent control with broadband down-converted light. *Phys Rev Lett* **93**, 023005 (2004).
10. Abram, I., Raj, R. K., Oudar, J. L. & Dolique, G. Direct Observation of the Second-Order Coherence of Parametrically Generated Light. *Phys Rev Lett* **57**, 2516 (1986).
11. Pe'er, A., Dayan, B., Friesem, A. A. & Silberberg, Y. Temporal shaping of entangled photons. *Phys Rev Lett* **94**, 073601 (2005).

12. Landes, T. *et al.* Experimental feasibility of molecular two-photon absorption with isolated time-frequency-entangled photon pairs. *Physical Review Research,* **3**, 33154 (2021).
13. Tabakaev, D. *et al.* Spatial Properties of Entangled Two-Photon Absorption. *Phys Rev Lett* **129**, 183601 (2022).
14. Landes, T., Smith, B. J. & Raymer, M. G. Limitations in Fluorescence-Detected Entangled Two-Photon-Absorption Experiments: Exploring the Low- to High-Gain Squeezing Regimes. (2024).
15. Villabona-Monsalve, J. P., Calderón-Losada, O., Nuñez Portela, M. & Valencia, A. Entangled Two Photon Absorption Cross Section on the 808 nm Region for the Common Dyes Zinc Tetraphenylporphyrin and Rhodamine B. *Journal of Physical Chemistry A* **121**, 7869–7875 (2017).
16. Lee, D. I. & Goodson, T. Entangled photon absorption in an organic porphyrin dendrimer. *Journal of Physical Chemistry B* **110**, 25582–25585 (2006).
17. Tabakaev, D. *et al.* Energy-time-entangled two-photon molecular absorption. *Phys Rev A (Coll Park)* **103**, 033701 (2021).
18. Varnavski, O. *et al.* Colors of entangled two-photon absorption. *Proc Natl Acad Sci U S A* **120**, e2307719120 (2023).
19. Mikhaylov, A. *et al.* Hot-Band Absorption Can Mimic Entangled Two-Photon Absorption. *Journal of Physical Chemistry Letters* **13**, 1489–1493 (2022).
20. Hickam, B. P., He, M., Harper, N., Szoke, S. & Cushing, S. K. Single-Photon Scattering Can Account for the Discrepancies among Entangled Two-Photon Measurement Techniques. *Journal of Physical Chemistry Letters* **13**, 4934–4940 (2022).
21. Corona-Aquino, S. *et al.* Experimental Study of the Validity of Entangled Two-Photon Absorption Measurements in Organic Compounds. *Journal of Physical Chemistry A* **126**, 2185–2195 (2022).
22. Triana-Arango, F., Ramírez-Alarcón, R. & Ramos-Ortiz, G. Entangled Two-Photon Absorption in Transmission-Based Experiments: Deleterious Effects from Linear Optical Losses. *Journal of Physical Chemistry A* **128**, 2210–2219 (2024).
23. Triana-Arango, F., Ramos-Ortiz, G. & Ramírez-Alarcón, R. Spectral Considerations of Entangled Two-Photon Absorption Effects in Hong-Ou-Mandel Interference Experiments. *Journal of Physical Chemistry A* **127**, 2608–2617 (2023).
24. Javanainen, J. & Gould, P. L. Linear intensity dependence of a two-photon transition rate. *Phys Rev A (Coll Park)* **41**, 5088 (1990).
25. Gea-Banacloche, J. Two-photon absorption of nonclassical light. *Phys Rev Lett* **62**, 1603 (1989).
26. Raymer, M., Landes, T. Theory of two-photon absorption with broadband squeezed vacuum. *Physical Review A,* **106**, 13717 (2022).
27. Chekhova, M. V., Leuchs, G. & Zukowski, M. Bright squeezed vacuum: Entanglement of macroscopic light beams. *Opt Commun* **337**, 27–43 (2015).
28. Saleh, B. E. A., Jost, B. M., Fei, H. B. & Teich, M. C. Entangled-Photon Virtual-State Spectroscopy. *Phys Rev Lett* **80**, 3483 (1998).
29. Pawlicki, M., Collins, H. A., Denning, R. G. & Anderson, H. L. Two-Photon Absorption and the Design of Two-Photon Dyes. *Angewandte Chemie International Edition* **48**, 3244–3266 (2009).
30. Gäbler, T. B., Hendra, P., Jain, N. & Gräfe, M. Photon Pair Source based on PPLN-Waveguides for Entangled Two-Photon Absorption. *Advanced Physics Research* **3**, 2300037 (2024).

31. Cameron, P., Courme, B., Faccio, D. & Defienne, H. Shaping the spatial correlations of entangled photon pairs. *Journal of Physics: Photonics* **6**, 033001 (2024).
32. Scott, R. *et al.* Two Photon Absorption in II-VI Semiconductors: The Influence of Dimensionality and Size. *Nano Lett* **15**, 4985–4992 (2015).
33. Xu, C. & Webb, W. W. Measurement of two-photon excitation cross sections of molecular fluorophores with data from 690 to 1050 nm. *J. Opt. Soc. Am. B* **13**, (1996).
34. Gonçalves, I. M. *et al.* Nonlinear Optical Properties of 2D CdSe Nanoplatelets in a Nonresonant Regime. *Journal of Physical Chemistry C* **127**, 16679–16686 (2023).
35. Li, B. & Hofmann, H. F. Enhancement of broadband entangled two-photon absorption by resonant spectral phase flips. *Phys Rev A* **108**, 013706 (2023).
36. Abouraddy, A. F., Saleh, B. E. A., Sergienko, A. V. & Teich, M. C. Role of Entanglement in Two-Photon Imaging. *Phys Rev Lett* **87**, 123602 (2001).
37. Li, H., Piryatinski, A., Srimath Kandada, A. R., Silva, C. & Bittner, E. R. Photon entanglement entropy as a probe of many-body correlations and fluctuations. *Journal of Chemical Physics* **150**, 184106 (2019).
38. D'Angelo, M., Chekhova, M. V. & Shih, Y. Two-Photon Diffraction and Quantum Lithography. *Phys Rev Lett* **87**, 013602 (2001).
39. Chen, Y., Zopf, M., Keil, R., Ding, F. & Schmidt, O. G. Highly-efficient extraction of entangled photons from quantum dots using a broadband optical antenna. *Nature Communications 2018 9:1* **9**, 1–7 (2018).
40. Fedorov, M. V., Mikhailova, Y. M. & Volkov, P. A. Gaussian modelling and Schmidt modes of SPDC biphoton states. *Journal of Physics B: Atomic, Molecular and Optical Physics* **42**, 175503 (2009).
41. Schneeloch, J. & Howell, J. C. Introduction to the transverse spatial correlations in spontaneous parametric down-conversion through the biphoton birth zone. *Journal of Optics* **18**, 053501 (2016).
42. Defienne, H., Reichert, M. & Fleischer, J. W. General Model of Photon-Pair Detection with an Image Sensor. *Phys Rev Lett* **120**, 203604 (2018).
43. Reichert, M., Defienne, H. & Fleischer, J. W. Massively Parallel Coincidence Counting of High-Dimensional Entangled States. *Scientific Reports 2018 8:1* **8**, 1–7 (2018).
44. Saleh, B. E. A., Teich, M. C., Abouraddy, A. F. & Sergienko, A. V. Entangled-photon Fourier optics. *JOSA B, 1174-1184* **19**, 1174–1184 (2002).
45. Defienne, H., Reichert, M., Fleischer, J. W. & Faccio, D. Quantum image distillation. *Sci Adv* **5**, (2019).
46. Courme, B., Cameron, P., Faccio, D., Gigan, S. & Defienne, H. Manipulation and Certification of High-Dimensional Entanglement through a Scattering Medium. *PRX Quantum* **4**, 010308 (2023).
47. Ithurria, S. & Talapin, D. V. Colloidal Atomic Layer Deposition (c-ALD) using self-limiting reactions at nanocrystal surface coupled to phase transfer between polar and nonpolar media. *J Am Chem Soc* **134**, 18585–18590 (2012).

**Acknowledgements**

R.P. and H.D. thank Dan Oron (Weizmann Institute) and Brian J. Smith (University of Oregon) for critical reading of the manuscript and advice. R.P. and H.D. also thank Daniele Faccio (University of Glasgow) for advice on experiments. R.P. thanks Clare College, University of Cambridge, for funding the work via a Junior Research Fellowship. H.D. acknowledges funding from an ERC Starting Grant (grant no. SQIMIC-101039375). P C and H D acknowledge support from SPIE Early Career Researcher Accelerator fund in Quantum Photonics.

**Author Contributions**
R.P. and H.D. designed the project. R.P. performed built and performed the experiments, and analyzed the data with guidance and help from P.C., C.V. and B.C.. S.I. and E.L. prepared the CdSe nanoplatelets. A.W.C. provided insightful discussions on the results. H.D. supervised the work. All author contributed to writing of the manuscript.